\documentclass[12pt]{article}
\usepackage{mathrsfs}
\usepackage{amscd}
\usepackage{amsfonts}
\usepackage{CJK}
\usepackage{amssymb}
\usepackage{amsmath}
\usepackage{color}
\usepackage{cite}
\usepackage{bbm}
\usepackage{graphicx}
\usepackage{setspace}

\newtheorem{theorem}{Theorem}[section]

\newtheorem{cor}[theorem]{Corollary}

\newtheorem{define}[theorem]{Definition}
\newtheorem{remark}[theorem]{Remark}

\newcommand\btd{\raise 2pt \hbox{$\hat\bigtriangledown$}\hskip 1.5pt}
\newcommand\bt{\raise 2pt \hbox{$\bigtriangledown$}\hskip 1.5pt}

\def\no{\nonumber}

\begin{document}
\title{Conservation laws of partial differential equations: symmetry, adjoint symmetry and nonlinear self-adjointness}
\author{Zhi-Yong Zhang \footnote{E-mail: zhiyong-2008@163.com; Tel:+86 010 88803103 } \ \ \ \  \ \ \ \ 
\\ \small~College of Sciences, North China University of
Technology, Beijing 100144, P.R. China}
\date{}
\maketitle

\noindent{\bf Abstract:} 
Nonlinear self-adjointness method for constructing conservation laws of partial differential equations (PDEs) is further studied.
We show that any adjoint symmetry of PDEs is a
differential substitution of nonlinear self-adjointness and vice
versa. Consequently, each symmetry of PDEs corresponds to a
conservation law via a formula if the system of PDEs is nonlinearly
self-adjoint with differential substitution.
As a byproduct, we find that the set of
differential substitutions includes the set of conservation law
multipliers as a subset. The results are illustrated by
three typical examples.

\noindent{\bf Keywords:} Nonlinear self-adjointness with
differential substitution, Adjoint symmetry, Conservation law,
Multiplier

\section{Introduction}
Conservation laws describe physical properties of the PDEs modeling
phenomena. They are used for the study of PDEs such as detecting integrability
and linearization, determining constants of motion, finding potentials and constructing nonlocally-related systems,
checking accuracy of numerical solution methods \cite{Olver,blu11}.

It is well-known that Noether' theorem established a close
connection between symmetries and conservation laws for the PDEs
possessing a variational structure \cite{Olver,blu11}.
However, application of Noether' approach relies on the
following two conditions which heavily hinder the construction of
conservation laws in such way:

(1). The PDEs under consideration must be derived from a variational
principle, i.e. they are Euler-Lagrange equations.

(2). The used symmetries must leave the variational integral
invariant, which means that not each symmetry of the PDEs can
generate a conservation law via Noether' theorem. Note that the
symmetry stated here and below refers to the generalized symmetry of
PDEs if no special notations are added.

Thus many researchers dedicated to develop new approaches to get around the
limitations of Noether's theorem
\cite{lma-1979,kara-2006,sb-1997,sb-2002a,sb-2002b}. 
In particular, multiplier method is very effective to construct conservation laws no matter
whether or not the PDEs admit a variational principle.
Olver's use of the Euler operator provides a feasible way to find all multipliers in principle \cite{Olver} while an algorithmic version of this method is the direct construction method where the corresponding local conservation laws are presented through an homotopy integral
formula \cite{sb-1997,sb-2002a,sb-2002b}.

Recently, Ibragimov provides a special method, named by nonlinear self-adjointness method, to construct some conservation laws of PDEs
\cite{ib-2011,tt-2012,Fs-2012}. The two required
conditions of this approach are the admitted symmetries
and the differential substitutions which convert nonlocal
conservation laws to local ones. As for the first requirement,
finding the symmetries of the PDEs, there exist a number of
well-developed methods and computer algebra programs
\cite{nj-2004,ch-2007}. However, the way to obtain the
required differential substitutions is only to use the equivalent
identity of the definition involving complicated computations, which
even makes us cannot get the expected results
\cite{ib-2011,zhang-2015,gan-2014}.

Therefore in this paper, we show the following two main results:

1. We show that each adjoint symmetry of the PDEs is a differential
substitution and vice versa, which gives a positive answer for
finding the differential substitutions with a new way. As a
byproduct, we find that the set of differential substitutions
contains the one of multipliers as a subset.

2. A direct connection among the symmetry, adjoint symmetry
and conservation law of the PDEs is expressed by an explicit formula, where the
formula only involves differential operation instead of integral
operation and thus can be fully implemented on a computer.
The above results are exemplified by three illustrated PDEs.

It should be noted that multiplier method does not require the
symmetry information of
PDEs but connected with the symmetry and adjoint symmetry \cite{sb-1997,sb-2002a,sb-2002b}.
On the solution space of
the given system of PDEs, multipliers are symmetries provided that
its linearized system is self-adjoint, otherwise they are adjoint
symmetries and can be obtained by choosing from the set of adjoint
symmetries by virtue of the so-called adjoint invariance conditions \cite{sb-2002a,sb-2002b}.
Quite recently, Anco shows that the general conservation law formula by Ibragimov is equivalent to a standard formula for the action of an infinitesimal symmetry on a conservation law \cite{anco-2016a,anco-2016b,anco-2016c}, what is more, the formula and its earlier version cannot in general produce all admitted conservation laws which are illustrated by some explicit examples \cite{anco-symm}.

The remainder of the paper is arranged as follows. In Section 2,
some related notions and principles are reviewed and the main
results are given. In Section 3, three different PDEs are
considered to illustrate the connections among symmetry, adjoint symmetry and the differential substitution of nonlinear self-adjointness of PDEs.
The last section contains a conclusion of the results.
\section{Main results}
In this section, we first review some related notions and
principles, and then give the main results of the paper.
\subsection{Preliminaries}
\subsubsection{Symmetry, adjoint symmetry and conservation law}
Consider a system of $m$ PDEs with $r$th-order
\begin{eqnarray}\label{perturb}
E^{\alpha}(x,u,u_{(1)},\cdots,u_{(r)})=0, ~~~~\alpha=1,2,\dots,m,
\end{eqnarray}
where $x=(x^1,\dots,x^n)$ is an independent variable set and
$u=(u^1,\dots,u^m)$ is a dependent variable set, $u_{(i)}$ denotes
all $i$-th $x$ derivatives of $u$. System (\ref{perturb}) is normal if each PDE is expressed in a solved form for some leading
derivative of $u$ such that all other terms in the system contain neither the leading derivative
nor its differential consequences \cite{sb-2002a,sb-2002b}.

On the solution space of the given PDEs, a symmetry is determined by
its linearized system while the adjoint symmetry is defined as the
solution of the adjoint of the linearized system \cite{blu11,Olver}.

In particular, the determining system of a symmetry 
$X_\eta=\eta^i(x,u,u_{(1)},\dots,u_{(s)})\partial_{u^i}$ is the
linearization of system (\ref{perturb}) annihilating on its solution
space, that is,
\begin{eqnarray}\label{sym-det}
&&(\mathscr{L}_E)^\alpha_\rho \eta^\rho=\frac{\partial
E^\alpha}{\partial u^\rho}\eta^\rho+\frac{\partial
E^\alpha}{\partial
u^\rho_{i_1}}D_{i_1}\eta^\rho+\dots+\frac{\partial
E^\alpha}{\partial u^\rho_{i_1\dots i_r}}D_{i_1}\dots
D_{i_r}\eta^\rho=0
\end{eqnarray}
holds for all solutions of system  (\ref{perturb}). The $m$-tuple
$\eta=(\eta^1,\eta^2,\dots,\eta^m)$ is called the characteristic of
the symmetry. In (\ref{sym-det}) and below, the summation
convention for repeated indices will be used and $D_i$ denotes the total
derivative operator with respect to $x^i$,
\begin{eqnarray}
&&\no D_i = \frac{\partial}{\partial x^i} +u_{i}^\sigma
\frac{\partial}{\partial u^\sigma}
+u_{ij}^\sigma\frac{\partial}{\partial u_{j}^\sigma} +\dots,
~~~i=1,2,\dots,n.
\end{eqnarray}
The adjoint equations of system
(\ref{sym-det}) are
\begin{eqnarray}\label{adsym-det}
(\mathscr{L}_E^*)_\alpha^\rho \omega_\rho=\omega_\rho \frac{\partial
E^\rho}{\partial u^\alpha}-D_{i_1}\bigg(\omega_\rho\frac{\partial
E^\rho}{\partial u^\alpha_{i_1}}\bigg)+\dots+(-1)^r D_{i_1}\dots
D_{i_r}\bigg(\omega_\rho\frac{\partial E^\rho}{\partial
u^\alpha_{i_1\dots i_r}}\bigg)=0,
\end{eqnarray}
which are the determining equations for an adjoint symmetry
$X_{\omega}=\omega_\rho(x,u,u_{(1)},\cdots,u_{(r)})\partial_{u^\rho}$
of system (\ref{perturb}).

In general, solutions of the adjoint symmetry determining system
(\ref{adsym-det}) are not solutions of the symmetry determining
system (\ref{sym-det}). However, if the linearized system
(\ref{sym-det}) is self-adjoint, then adjoint symmetries are
symmetries and system (\ref{perturb}) has a variational principle
and thus Noether' approach is applicable in this case \cite{Olver}.

\begin{define} (Conservation law \cite{Olver})
A conservation law of system (\ref{perturb}) is a divergence
expression
\begin{equation}
\no D_i(C^i)=D_1(C^1)+\dots+D_n(C^n)=0
\end{equation}
for all solutions of system (\ref{perturb}).
If for some $j=1,\dots,n$, $x^j=t$, then $C^t$ is called the conserved density and the other $C^i(i\neq j)$ are called
the spatial fluxes and the pair $(C^t,C^i)$ is called a conserved current. \end{define}

A conservation law is trivial if for all solutions of system (\ref{perturb}),
$C^i = D_k \Theta^{ik}$ with $\Theta^{ik}=-\Theta^{ki}$  for some expressions $\Theta^{ik}=\Theta^{ik}(x,u,u_{(1)},\cdots,u_{(r-1)})$. 
Any two equivalent conservation laws differ by a trivial conservation law.  For a given PDEs, the set of all nontrivial conservation laws (up to equivalence) forms a vector space.
\subsubsection{Nonlinear self-adjointness with differential substitution}

We begin with nonlinear self-adjointness introduced by Ibragimov
\cite{ib-2011}, whose main idea is first to turn the system of PDEs
into Lagrangian equations by artificially adding new variables, and
then to apply the theorem proved in \cite{nh-2007} to construct
local and nonlocal conservation laws.

Specifically, let $\mathcal {L}$ be the formal Lagrangian of system
(\ref{perturb}) written as
\begin{eqnarray}\label{lagrangian}
&& \mathcal {L} = v^{\beta}E^{\beta}(x,u,u_{(1)},\dots,u_{(r)}),
\end{eqnarray}
where $ v^{\beta}$ are new introduced dependent variables, then
the adjoint equations of system (\ref{perturb}) are defined by
\begin{eqnarray}\label{adequation}
(E^{\alpha})^{\ast}(x,u,v,u_{(1)},v_{(1)},\cdots,u_{(r)},v_{(r)})=\frac{\delta\mathcal
{L}}{\delta u^{\alpha}}=0,
\end{eqnarray}
where $v=(v^1,\dots,v^m)$ and hereinafter, $\delta/\delta u^{\alpha}$
is the Euler operator
 \begin{eqnarray}\label{var-op}
\frac{\delta}{\delta u^{\alpha}}=\frac{\partial}{\partial
u^{\alpha}}+\sum_{s=1}^{\infty}(-1)^sD_{i_1}\dots
D_{i_s}\frac{\partial}{\partial u^{\alpha}_{i_1\dots i_s}}.
\end{eqnarray}
Then the definition of nonlinear self-adjointness of system
(\ref{perturb}) is given as follows.

\begin{define} \label{def-non}(Nonlinear self-adjointness \cite{ib-2011}) The
system (\ref{perturb}) is said to be nonlinearly self-adjoint if the
adjoint system (\ref{adequation}) is satisfied for all solutions
of system (\ref{perturb}) upon a substitution $v=\varphi(x,u)$ such
that $\varphi(x,u)\neq 0$.\end{define}

Here, $\varphi(x,u)=(\varphi^1(x,u),\dots,\varphi^m(x,u))$ and
$v=\varphi(x,u)$ means $v^i=\varphi^i(x,u)$, $\varphi(x,u)\neq 0$
means that not all elements of $\varphi(x,u)$ equal zero and is
called a nontrivial substitution. Definition \ref{def-non} is
equivalent to the following identities holding for the undetermined non-singular
functions
$\lambda_{\alpha}^{\beta}=\lambda_{\alpha}^{\beta}(x,u,u_{(1)}\dots,u_{(r)})$
\begin{eqnarray}\label{equ-id1}
&&
(E^{\alpha})^{\ast}(x,u,v,u_{(1)},v_{(1)},\cdots,u_{(r)},v_{(r)})_{|_{v=\varphi}}
=\lambda_{\alpha}^{\beta}E^{\beta},
\end{eqnarray}
which is applicable in the proofs and computations.

As an extension of the substitution, if
$v=\varphi(x,u,u_{(1)},\dots,u_{(s)})$, then it is called nonlinear
self-adjointness with differential substitution \cite{ib-2011,gan-2014,zhang-2015}. 
\begin{define} \label{def-ger}(Nonlinear self-adjointness with differential
substitution) The system (\ref{perturb}) is said to be nonlinearly
self-adjoint with differential substitution if the adjoint system
(\ref{adequation}) is satisfied for all solutions of system
(\ref{perturb}) upon a substitution
$v=\varphi(x,u,u_{(1)},\dots,u_{(s)})$ such that $v\neq 0$.
\end{define}

Similarly, Definition \ref{def-ger} is equivalent to the following
equality
\begin{eqnarray}\label{equ-id1-id2}
&&\no(E^{\alpha})^{\ast}(x,u,v,u_{(1)},v_{(1)},\cdots,u_{(r)},v_{(r)})_{|_{v=\varphi(x,u,u_{(1)},\dots,u_{(s)})}}\\
&&\hspace{5cm}=(\lambda_{\alpha}^{\beta}+\lambda_{\alpha}^{\beta
i_1}D_{i_1}+\dots+\lambda_{\alpha}^{\beta i_1\dots i_s}D_{i_1}\dots
D_{i_s})E^\beta,
\end{eqnarray}
where $\lambda_{\alpha}^{\beta},\lambda_{\alpha}^{\beta
i_1},\dots,\lambda_{\alpha}^{\beta i_1\dots i_s}$ are undetermined
functions of arguments $x,u,u_{(1)},\dots, u_{(r+s)}$  and  non-singular on the solutions of the given PDE system (\ref{perturb}) respectively.
Since the highest order derivatives in
$\lambda_{\alpha}^{\beta},\lambda_{\alpha}^{\beta
i_1},\dots,\lambda_{\alpha}^{\beta i_1\dots i_s}$ may be higher than
the highest order derivative in $E^\alpha$, the right side of system
(\ref{equ-id1-id2}) may not linear in $D_{i_1}\dots
D_{i_k}E^\alpha,$ $k=1,\dots,m$, and thus application of equality
(\ref{equ-id1-id2}) to find the differential substitutions is a
difficult task \cite{zhang-2015}. For example, when using equality
(\ref{equ-id1-id2}) to search for the differential substitutions of
Klein-Gordon equation (\ref{k-g-eq}) studied in Subsection 3.3,
equality (\ref{equ-id1-id2}) becomes Eq.(\ref{det}) which is not
linear in $D_x^k G$. Therefore, it is necessary to develop new
approaches to search for differential substitution.

After finding the differential substitutions of nonlinear
self-adjointness, we will use the following theorem to construct
conservation laws of the system \cite{nh-2007}.

\begin{theorem} \label{con-law}
Any infinitesimal symmetry (Local and nonlocal)
\begin{eqnarray}
&&\no X=\xi^i(x,u,u_{(1)},\dots)\frac{\partial}{\partial
x^i}+\eta^{\sigma}(x,u,u_{(1)},\dots)\frac{\partial}{\partial
u^{\sigma}}
\end{eqnarray}
of system (\ref{perturb}) leads to a conservation law $D_i(C^i)_{|E^\alpha=0}=0$
constructed by the formula
\begin{eqnarray}\label{formula}
&&\no C^i=\xi^i\mathcal {L}+W^{\sigma}\bigg[\frac{\partial \mathcal
{L}}{\partial u_i^{\sigma}}-D_j(\frac{\partial \mathcal
{L}}{\partial u_{ij}^{\sigma}})+D_jD_k(\frac{\partial \mathcal
{L}}{\partial u_{ijk}^{\sigma}})-\dots\bigg]\\&&\hspace{0.8cm}
+D_j(W^{\sigma}) \bigg[\frac{\partial \mathcal {L}}{\partial
u_{ij}^{\sigma}}-D_k(\frac{\partial \mathcal {L}}{\partial
u_{ijk}^{\sigma}})+\dots\bigg]+D_jD_k(W^{\sigma})\bigg[\frac{\partial
\mathcal {L}}{\partial u_{ijk}^{\sigma}}-\dots\bigg]+\dots,
\end{eqnarray}
where $W^{\sigma}=\eta^{\sigma}-\xi^ju_j^{\sigma}$ and $\mathcal
{L}$ is the formal Lagrangian (\ref{lagrangian}) which is written in
the symmetric form about the mixed derivatives. \end{theorem}

Note that the first term $\xi^i\mathcal {L}$ in the right side of formula (\ref{formula}) is actually trivial since $\mathcal {L}=0$ holds identically for any solution of the given PDE system (\ref{perturb}) \cite{ib-2011,anco-symm}.
\subsubsection{Multiplier}

Multipliers are a set of functions which multiplies a system of PDEs
in order to make the system get a divergence form, then for any
solution of the equations this divergence will equal zero and one
will get a conservation law. 
\begin{define}\label{df-mul}(Multiplier \cite{sb-2002b})
A multiplier for system (\ref{perturb}) is a set of non-singular functions on the solution space
\begin{equation}\label{mult}
\Lambda=\{\Lambda_1(x,u,u_{(1)},\dots,
u_{(s)}),\,\dots,\,\Lambda_m(x,u,u_{(1)},\dots,
u_{(s)})\},\end{equation} satisfying
\begin{equation}\label{mul}
\Lambda_\beta(x,u,u_{(1)},\dots,
u_{(s)})E^{\beta}(x,u,u_{(1)},\dots,u_{(r)})=D_i(C^i)
\end{equation}
with some expressions $C^i$ for any function $u$.
\end{define}

For a normal PDE system (\ref{perturb}) with no differential identities, Eq.(\ref{mul}) demonstrates that a conservation law is trivial
if the corresponding multiplier vanishes identically on the solution space of system (\ref{perturb}), otherwise it is nontrivial \cite{anco-2016b}.
Thus there is a one-to-one correspondence between conservation laws (up to equivalence) and multipliers evaluated
on the solution space of the normal system (\ref{perturb}) without differential identities.
Since Euler operator $\delta/\delta u^{\sigma}$ with
$\sigma=1,2,\dots,m$ acting on the divergence expression $D_i(C^i)$
yields zero identically, so the following theorem is established \cite{Olver,blu11}.
 \begin{theorem}
A non-singular local multiplier (\ref{mult}) yields a local
conservation law for the PDEs system (\ref{perturb}) if and only if
the set of identities
\begin{equation}\label{det-mult}
\frac{\delta}{\delta
u^{\sigma}}\big(\Lambda_\beta(x,u,u_{(1)},\dots,
u_{(s)})E^{\beta}(x,u,u_{(1)},\dots,u_{(r)})\big)=0
\end{equation}
holds for arbitrary functions $u=u(x)$.
\end{theorem}

Since system (\ref{det-mult}) holds for arbitrary $u=u(x)$, one can treat
each $u$  and its derivatives as independent variables, and
consequently separate system (\ref{det-mult}) into an
over-determined linear PDEs system about $\Lambda_\beta$ whose solutions are multipliers.
When the calculation works on the solution space of the given PDEs
expressed in a Cauchy-Kovalevskaya form, multipliers are
selected from the set of adjoint symmetries using the Helmholtz-type
conditions \cite{sb-2002a,sb-2002b}.
\subsection{Main results}
We first give an equivalent definition of nonlinear self-adjointness
with differential substitution. Definition \ref{def-ger} means that
adjoint system (\ref{adequation}), after inserted by the
differential substitution $v=\varphi(x,u,u_{(1)},\dots,u_{(s)})$,
holds identically on the solution space of original system
(\ref{perturb}). This property can be used as the following
alternative definition for nonlinear self-adjointness with
differential substitution.

\begin{define} \label{def-ger-eqv}(Nonlinear self-adjointness with differential
substitution) The system (\ref{perturb}) is nonlinearly self-adjoint
with differential substitution if the adjoint system
(\ref{adequation}) upon a nontrivial differential substitution
$v=\varphi(x,u,u_{(1)},\dots,u_{(s)})$ holds on the solution space
of system (\ref{perturb}).
\end{define}

In the sense of Definition \ref{def-ger-eqv}, nonlinear
self-adjointness with differential substitution is equivalent to the
following equality
\begin{eqnarray}\label{equ-id1-id1}
&&(E^{\alpha})^{\ast}_{|_{v=\varphi}}=\frac{\delta\mathcal
{L}}{\delta
u^{\alpha}}_{|_{v=\varphi}}=0,~~~\mbox{when}~E^{\alpha}=0,
\end{eqnarray}
which is called the determining system of differential substitution.

Following the idea of Definition \ref{def-ger-eqv}, we obtain the
following results. Note that though Theorem \ref{th-2} has been obtained in \cite{anco-2016a,anco-symm}, here we show it from the point of view of the equivalent Definition \ref{def-ger-eqv} of nonlinear self-adjointness.

\begin{theorem}\label{th-2}
Any adjoint symmetry of system (\ref{perturb}) is a differential
substitution of nonlinear self-adjointness and vice versa.
\end{theorem}

\emph{Proof.} We start with Eq.(\ref{equ-id1-id1}). Since $v$ is a
new introduced dependent variable set, then on the solution space of
system (\ref{perturb}), Eq.(\ref{equ-id1-id1}) can be explicitly
expressed as
\begin{eqnarray}\label{equ-id1-id3}
&&\no(E^{\alpha})^{\ast}_{|_{v=\varphi}}=\frac{\delta\mathcal
{L}}{\delta u^{\alpha}}_{|_{v=\varphi}}\\
&&\hspace{1.6cm}=\left[v^\beta\frac{\partial E^\beta}{\partial
u^{\alpha}}+\sum_{r=1}^{\infty}(-1)^rD_{i_1}\dots
D_{i_r}\left(v^\beta\frac{\partial E^\beta}{\partial
u^{\alpha}_{i_1\dots i_r}}\right)\right]_{|_{v=\varphi}}=0.
\end{eqnarray}

Obviously, system (\ref{equ-id1-id3}) and the adjoint symmetry
determining system (\ref{adsym-det}) are the same in the form, 
thus solutions of Eq.(\ref{equ-id1-id3}) satisfy
Eq.(\ref{adsym-det}) and vice versa. The proof ends. $\square$

Theorem \ref{th-2} provides an effective way to search for differential
substitution of nonlinear self-adjointness, which is equivalent to
find the adjoint symmetry of PDEs. Furthermore, for a given differential
substitution of nonlinear self-adjointness, formula (\ref{formula})
can generate a conservation law with the symmetry of system
(\ref{perturb}), thus together with Theorem \ref{con-law} and
Theorem \ref{th-2}, we formulate the following algorithm for constructing conservation laws of PDEs.

Step 1: Compute symmetries and adjoint symmetries admitted by the
 PDEs.

Step 2: Construct the formal Lagrangian $\mathcal {L}$ and find the
differential substitutions.

By Theorem \ref{th-2}, the admitted adjoint symmetries are the required differential substitutions of nonlinear self-adjointness.

Step 3: With the above known symmetry information, use formula (\ref{formula}) to
construct conservation laws of the PDEs.

 Since computing symmetry and adjoint
symmetry is an algorithmic procedure, thus a variety of symbolic
manipulation programs have been developed for many computer algebra
systems (See \cite{nj-2004,ch-2007} and references therein).
Furthermore, the general conservation law formula (\ref{formula})
only involve differential operation instead of integral operation.
Hence, the proposed algorithm can be fully implemented on a
computer.

\begin{remark}
For the PDEs having a Lagrangian, nonlinear self-adjointness method has two merits in comparison with Noether' theorem: neither constructing a Lagrangian nor choosing the
variational symmetries from the set of admitted symmetries.
\end{remark}

However, it should be noted that the constructed conservation laws by the Ibragimov's method is incomplete and may be trivial, thus one should adopt some tools to check the triviality such as the physical properties of conservation laws or whether the obtained conservation law corresponds to some nontrivial multiplier \cite{anco-symm}.

To end this section, we study the connection between nonlinear
self-adjointness with differential substitution and multiplier
method. Multiplier method for the normal PDEs is further studied in
\cite{sb-2002a,sb-2002b,anco-2016a,anco-2016b}, which states that multipliers can be
obtained by choosing from the set of adjoint symmetries with the
 adjoint invariance conditions, thus by Theorem
\ref{th-2}, we have:

\begin{cor}\label{cor-1}
For the normal PDE system (\ref{perturb}), the set of
differential substitutions contains the one of multipliers as a
subset.
\end{cor}

It is well-known that any non-variational symmetry of an Euler-Lagrange system is an adjoint symmetry (which coincides with a symmetry) but not a multiplier \cite{blu11}, thus from Corollary \ref{cor-1}, there exist some adjoint
symmetries which are differential substitutions but not multipliers
of system (\ref{perturb}), this case will be exemplified by a
nonlinear wave equation in the next section. Another simple observation is that if a PDE system admits a conservation law, then the multiplier is an adjoint-symmetry and hence the system is nonlinearly self-adjoint. 

\section{Three illustrated examples}
In this section, we consider three examples, where the first example
is a nonlinear wave equation used to demonstrate the result of
Corollary \ref{cor-1}, the second one is the Thomas equation which
shows that nonlinear self-adjointness with differential substitution
method can deal with the PDEs without having a Lagrangian and finds
new substitutions, and the third one is the Klein-Gordon equation
used to illustrate the effectiveness of nonlinear self-adjointness
with differential substitution when dealing with the PDEs derived
from a variational principle. Note that in this section $u=u(x,t)$
is a dependent variable of two independent variables $x$ and $t$, and we will not differentiate $u_{x}$ and
$\partial_xu$, which is also suitable for the cases of higher-order
derivatives.

Before going further, we first define two operators in order to
simplify some expressions in the computations. The symbol
$\mathscr{R}_\Delta[\partial^i_x\partial^j_tu]$ stands for
\begin{eqnarray}\label{linear}
&&\no\mathscr{R}_\Delta
[\partial^i_x\partial^j_tu]\Theta=\Delta_{\partial^i_x\partial^j_tu}
\Theta+\Delta_{\partial^{i+1}_x\partial^j_tu}D_{x}\Theta+\Delta_{\partial^{i}_x\partial^{j+1}_tu}D_t\Theta\\
&&\hspace{2.8cm}+\Delta_{\partial^{i+2}_x\partial^j_tu}D_x^2\Theta+
\Delta_{\partial^{i+1}_x\partial^{j+1}_tu}D_{x}D_t\Theta+\Delta_{\partial^i_x\partial^{j+2}_tu}D^2_t\Theta+\dots,
\end{eqnarray}
while
\begin{eqnarray}
&&\no\mathscr{W}_\Delta
[\partial^i_x\partial^j_tu]\Theta=
\Theta\Delta_{\partial^i_x\partial_t^ju}-\mathscr{D}_x\big(\Theta\Delta_{\partial^{i+1}_x\partial_t^ju}\big)-\mathscr{D}_t\big(\Theta\Delta_{\partial^i_x\partial_t^{j+1}u}\big)
\\
&&\no\hspace{3.2cm}
+\mathscr{D}_x^2\big(\Theta\Delta_{\partial^{i+2}_x\partial_t^ju}\big)
+\mathscr{D}_x\mathscr{D}_t\big(\Theta\Delta_{\partial^{i+1}_x\partial_t^{j+1}u}\big)
+\mathscr{D}_t^2\big(\Theta\Delta_{\partial^{i}_x\partial_t^{j+2}u}\big)+\dots,
\end{eqnarray}
where $i$ and $j$ are nonnegative integers, the symbols $\mathscr{D}_t$ and $\mathscr{D}_x$ are the total differential operators on the solution space of the corresponding targeted PDEs.

\subsection{A nonlinear wave equation}
The first example is to consider a nonlinear wave equation
\cite{blu11,sb-2002a}
\begin{eqnarray}\label{wave-speed}
&&E= u_{tt}-u^2u_{xx}-uu_{x}^2=0,
\end{eqnarray}
which has a variational principle given by the  action integral
$S=\int(u_t^2+u^2u_x^2)/2\,dtdx$ and thus the adjoint
symmetry and the symmetry are identical.

We first apply multiplier method to study conservation laws of
Eq.(\ref{wave-speed}). A function $\Lambda=\Lambda(x,t,u,u_x,u_t)$
is a multiplier of Eq.(\ref{wave-speed}) if and only if Euler
operator (\ref{var-op}) acting on the multiplication $\Lambda E$
yields zero for any $u=u(x,t)$, i.e.,
\begin{eqnarray}\label{wave-speed-det}
&&\no \frac{\delta (\Lambda E)}{\delta
u}=D_t^2\Lambda-u^2D_x^2\Lambda-2uu_xD_x\Lambda\\
&&\hspace{1.8cm}-(2uu_{xx}+u_x^2)\Lambda+E\Lambda_u-D_x(E\Lambda_{u_x})-D_t(E\Lambda_{u_t})=0.
\end{eqnarray}

Splitting Eq.(\ref{wave-speed-det}) with respect to $u_{tt}$ and its
differential results, we find that the determining system for
multiplier $\Lambda$ consists of the symmetry determining system
\begin{eqnarray}\label{wave-speed-det-1}
&&\mathscr{D}_t^2\Lambda-u^2D_x^2\Lambda-2uu_xD_x\Lambda-(2uu_{xx}+u_x^2)\Lambda=0,
\end{eqnarray}
where
$\mathscr{D}_t=\partial_t+u_t\partial_u+u_{xt}\partial_{u_x}+(u^2u_{xx}+uu_{x}^2)\partial_{u_t}
+\dots$ is the total derivative operator on the solution space of
Eq.(\ref{wave-speed}),  and
\begin{eqnarray}\label{wave-speed-det-2}
&& 2\Lambda_u-\mathscr{D}_t\Lambda_{u_t}-D_x\Lambda_{u_x}=0,
\end{eqnarray}
which is called the adjoint invariance condition. Note that $D_t$ is
connected with $\mathscr{D}_t$ by the equality $D_t\Lambda=\mathscr{D}_t\Lambda+\mathscr{R}_\Lambda[\partial_tu]E$ which implies
$D_t =\mathscr{D}_t$ on the solution space of  Eq.(\ref{wave-speed}).

On the other hand, Eq.(\ref{wave-speed}) is invariant under the
symmetry $X=(u-xu_x)\partial_u$, then function $\Lambda=u-xu_x$ is a
solution of Eq.(\ref{wave-speed-det-1}) but does not satisfy the
adjoint invariance condition (\ref{wave-speed-det-2}), thus it is
not a multiplier. However, by Theorem \ref{th-2}, function $\Lambda$
is a differential substitution of nonlinear self-adjointness for
Eq.(\ref{wave-speed}), then set the formal Lagrangian $$\mathcal
{L}=(u-xu_x)( u_{tt}-u^2u_{xx}-uu_{x}^2),$$ and by means of Theorem
\ref{con-law}, we obtain a nontrivial conservation law
$$D_tC_{(\ref{wave-speed})}^t+D_xC_{(\ref{wave-speed})}^x=\big[\mathscr{W}_{\mathscr{D}_t\eta}[\partial_tu](u-xu_x)
-\mathscr{W}_\eta[\partial_tu](u_t-xu_{xt})-2\eta-xD_x\eta\big]*E$$ given by
the formulae
\begin{eqnarray}
&&\no C_{(\ref{wave-speed})}^t=(u-xu_x)\mathscr{D}_t\eta-\eta
(u_t-xu_{xt}),\\\no
&&C_{(\ref{wave-speed})}^x=(xu_x-u)u^2D_x\eta-xu^2u_{xx}\eta,
\end{eqnarray}
where $X= \eta(x,t,u,u_x,u_t,\dots)\partial_u$ is a symmetry of
Eq.(\ref{wave-speed}).

For example, choose a simple translation symmetry of $t$ with characteristic $\eta=u_t$, then we obtain a nontrivial conservation law,
\begin{eqnarray}\label{con-wave}
&&D_tC_{(\ref{wave-speed})}^t+D_xC_{(\ref{wave-speed})}^x=-3u_t*E,
\end{eqnarray}
where
\begin{eqnarray}
&&\no C_{(\ref{wave-speed})}^t=(u-xu_x)(u^2u_{xx}+uu_{x}^2)- u_t^2+x u_t u_{xt},\\
&&\no C_{(\ref{wave-speed})}^x=x u^2 u_xu_{xt}-u^3u_{xt}-xu^2u_{xx}u_t.
\end{eqnarray}

In fact, multiplier $\Lambda=u_t$ generate a conservation law
\begin{eqnarray}\label{multi-wave}
&&D_t\big(-\frac{3}{2}u_t^2-\frac{3}{2}u^2u_x^2\big)+D_x(3u^2u_xu_t)=-3u_t*E,
\end{eqnarray}
which corresponds to conservation of energy.

It is well-known that for PDEs (\ref{wave-speed}) there is a one-to-one correspondence between conservation laws (up to equivalence) and multipliers evaluated on the solution space  \cite{anco-2016a,anco-2016b}, thus the conserved currents of conservation laws (\ref{con-wave}) and (\ref{multi-wave}) are connected by
\begin{eqnarray}
&&\no C_{(\ref{wave-speed})}^t=-\frac{3}{2}u_t^2-\frac{3}{2}u^2u_x^2+D_x\big(u^3u_x-\frac{1}{2}xu^2u_x^2+\frac{1}{2}xu_t^2\big),\\
&&\no C_{(\ref{wave-speed})}^x=3u^2u_xu_t+D_t\big(-u^3u_x+\frac{1}{2}xu^2u_x^2-\frac{1}{2}xu_t^2\big).
\end{eqnarray}
which means (\ref{con-wave}) and (\ref{multi-wave}) are equivalent by a trivial conservation law
$$D_x\Big[D_t\big(-u^3u_x+\frac{1}{2}xu^2u_x^2-\frac{1}{2}xu_t^2\big)\Big]+D_t\Big[D_x\big(u^3u_x-\frac{1}{2}xu^2u_x^2+\frac{1}{2}xu_t^2\big)\Big]\equiv0.$$
\subsection{The Thomas equation}
The Thomas equation is written as
\begin{eqnarray}\label{th-pde}
&&F=u_{xt}+\alpha u_x+\beta u_t+\gamma u_x u_t=0,
\end{eqnarray}
which arises in the study of chemical exchange process
\cite{rr-1966}, where the constants $\alpha,\beta$ and $\gamma$
satisfy $\alpha>0,\beta>0,\gamma\neq0$. The property of nonlinear
self-adjointness had been studied in \cite{ib-2011}. Note that
Eq.(\ref{th-pde}) is not variational due to the involved terms
$\alpha u_x$ and $\beta u_t$, but there exists a one-to-one correspondence between multipliers and conservation laws since it is a wave-type equation.

Following the infinitesimal symmetry criterion for PDEs
\cite{Olver}, the determining equation for a symmetry
$X=\eta(x,t,u,\partial_xu,\partial_tu,\dots)\partial_u$ of  Eq.(\ref{th-pde}) is
\begin{eqnarray}\label{th-det}
&& D_tD_x\eta+\alpha D_x\eta+\beta D_t\eta +\gamma u_xD_t\eta+\gamma
u_tD_x\eta=0
\end{eqnarray}
holding for all solutions of Eq.(\ref{th-pde}).
The adjoint equation of Eq.(\ref{th-det}) is
\begin{eqnarray}\label{th-pde-adj}
D_tD_x\psi-\alpha D_x\psi-\beta D_t\psi -\gamma u_xD_t\psi-\gamma
u_tD_x\psi+2\gamma (\alpha u_x+\beta u_t+\gamma u_x u_t)\psi=0,
\end{eqnarray}
which is the determining system of an adjoint symmetry
$X=\psi(x,t,u,\partial_xu,\partial_tu,\dots,)\partial_u$. Then by
Theorem \ref{th-2}, solutions of Eq.(\ref{th-pde-adj}) are the
differential substitution of nonlinear self-adjointness. Note that
the symmetry and adjoint symmetry do not contain $u_{xt}$ nor its
differential results since they can be expressed through
Eq.(\ref{th-pde}).

Introduce the formal Lagrangian of Eq.(\ref{th-pde}) in the
symmetric form for the mixed derivative $u_{xt}$ $$\mathcal
{L}=v\Big(\frac{1}{2}u_{xt}+\frac{1}{2}u_{tx}+\alpha u_x+\beta
u_t+\gamma u_x u_t\Big),$$ where $v$ is a new dependent variable,
then by formula (\ref{formula}), we obtain the following general
conservation law formulae.
\begin{theorem}
A conservation law $ \big(D_tC_{(\ref{th-pde})}^t+D_xC_{(\ref{th-pde})}^x\big)_{|F=0}=0$ of
Eq.(\ref{th-pde}) is given by
\begin{eqnarray}\label{con-law-1}
&&\no C_{(\ref{th-pde})}^t=(\gamma u_x v+\beta\, v)\eta+v \mathscr{D}_x\eta,\\
&&C_{(\ref{th-pde})}^x=(\gamma u_t v+\alpha\, v-\mathscr{D}_tv)\eta,
\end{eqnarray}
where differential substitution $v=\psi(x,t,u,\partial_xu,\partial_tu,\dots)$ determined by
Eq.(\ref{th-pde-adj}) and $X= \eta(x,t,u,u_x,u_t,\dots)\partial_u$
is a symmetry of Eq.(\ref{th-pde}). In (\ref{con-law-1}),  $\mathscr{D}_x$ and $\mathscr{D}_t$  are
the total derivative operators which expresses $u_{xt}$ and its
derivatives through Eq.(\ref{th-pde}).
\end{theorem}

In what follows, we first search for differential substitution and
then use formulae (\ref{con-law-1}) to construct conservation laws
of Eq.(\ref{th-pde}). Assume
$\psi=f(x,t,u)u_x+g(x,t,u)u_t+h(x,t,u)$, then substitute it into
Eq.(\ref{th-pde-adj}) and collect the coefficients of different
powers of $u_x,u_t,u_{xx}$ and $u_{tt}$, we obtain
\begin{eqnarray}\label{det-thomas}
&&\no h_{uu}-3 \gamma  h_u+2 \gamma ^2 h=0,\\
&&\no2 \alpha ^2 g-\alpha   g_t-2 \alpha  h_u-\gamma h_t+h_{tu}+2
\alpha  \gamma  h=0,\\\no &&2 \beta ^2 f-\beta f_x-2 \beta
h_u-\gamma  h_x+h_{xu}+2 \beta  \gamma h=0,\\
&& f_u-2 \gamma f= f_t-2 \alpha f=g_u-2 \gamma g= g_x-2 \beta g=0.
\end{eqnarray}
Solving system (\ref{det-thomas}) gives
\begin{eqnarray}\label{sub-to}
&&\no \hspace{-0.3cm}\psi=B(x,t)e^{\gamma u}+c_1e^{2(\gamma  u+\alpha  t+\beta x)}\\
&&\hspace{0.6cm}+e^{2(\gamma  u+\alpha  t+\beta x)}\Big[(c_3-c_2 t)
u_t+(c_2 x+c_4) u_x+\frac{1}{\gamma }( c_2\beta x- c_2\alpha t+
c_3\alpha+ c_4\beta)\Big],
\end{eqnarray}
where $c_1,\dots,c_4$ are arbitrary constants and $B(x,t)$ satisfies
$B_{xt}-\alpha B_x-\beta B_t=0$ such that $\psi\neq 0$.

Obviously, when $c_2=c_3=c_4=0$, adjoint symmetry (\ref{sub-to})
becomes the substitution of nonlinear self-adjointness, which is
identical to the results in \cite{ib-2011}, while expression
(\ref{sub-to}) with $B(x,t)=c_1=0$ is a new differential
substitution and may generate new nontrivial conservation laws of Eq.(\ref{th-pde}) .

\textbf{Example 1.} The first example is to consider the case $v=e^{2(\gamma u+\alpha t+\beta
x)}(u_t+\alpha/\gamma)$ and $\eta=-u_x$, then by  (\ref{con-law-1}) we have
\begin{eqnarray}
&&\no C_{(\ref{th-pde})}^t=-e^{2 (\gamma  u+\alpha  t+\beta x)}
\left[\alpha \beta u_{x}+\alpha \gamma u_{x}^2+\alpha u_{xx}+\gamma (\beta
   u_{x}+\gamma  u_{x}^2+u_{xx}) u_t\right],\\
&&\no C_{(\ref{th-pde})}^x=e^{2 (\gamma  u+\alpha  t+\beta  x)}
\left[ \gamma u_{x} u_{tt}+\alpha ^2u_{x}+2\alpha \gamma
   u_{x}u_t +\gamma^2  u_{x}u_t^2 \right],
\end{eqnarray}
which gives a conservation law
$$D_tC_{(\ref{th-pde})}^t+D_xC_{(\ref{th-pde})}^x=2\beta e^{2 (\gamma  u+\alpha  t+\beta  x)} \big(\alpha+\gamma u_t\big)*F.$$

\textbf{Example 2.} The second example is $v=e^{2(\gamma u+\alpha
t+\beta x)}(u_x+\beta/\gamma)$ and $\eta=f(x,t)e^{-\gamma u}$, where
$f$ satisfies $f_{xt}+\alpha f_x+\beta f_t=0$, then one has
\begin{eqnarray}
&&\no C_{(\ref{th-pde})}^t=e^{\gamma  u+2\alpha  t+2\beta  x}
\left(\gamma  f_x u_x+\beta  f_x+\beta  \gamma  f u_x+\beta ^2   f\right),~~~C_{(\ref{th-pde})}^x=\alpha\beta f e^{\gamma  u+2\alpha  t+2\beta  x},
\end{eqnarray}
which gives a conservation law in the form
\begin{equation}
\no D_tC_{(\ref{th-pde})}^t+D_xC_{(\ref{th-pde})}^x= e^{\gamma  u+2 \alpha  t+2 \beta  x}\left(f_x+\beta  f\right)*F.\end{equation}

\textbf{Example 3.} The last example is $v=e^{2(\gamma  u+\alpha
t+\beta x)}(xu_x-tu_t+(\beta x-\alpha t)/\gamma)$ and $\eta=-u_t$,
then a conservation law
\begin{eqnarray}
&&\no D_tC_{(\ref{th-pde})}^t+D_xC_{(\ref{th-pde})}^x=
e^{2 (\gamma  u+\alpha  t+\beta  x)} \big[ \alpha  \left(2 \alpha  t-2 \beta  x-2 \gamma  x u_x+1\right)+(\gamma +2 \alpha  \gamma  t) u_t\big]*F
   \end{eqnarray}
    is given by
\begin{eqnarray}
&&\no C_{(\ref{th-pde})}^t= e^{2 (\gamma  u+\alpha  t+\beta  x)}
\left(\alpha^2 t-\alpha\beta x+\alpha\gamma t u_t-\alpha\gamma x u_x\right) u_x,\\\no
&&C_{(\ref{th-pde})}^x=e^{2 (\gamma  u+\alpha  t+\beta  x)}\left(\alpha+\alpha^2t-\alpha\beta x+\gamma u_t+2\alpha\gamma t u_t+\gamma^2 t u_t^2+\gamma t u_{tt}\right)u_t.
\end{eqnarray}
\subsection{The Klein-Gordon equation}
In this section, we study the Klein-Gordon equation
\begin{equation}\label{k-g-eq}
G=u_{tt}-u_{xx}-g(u)=0,
\end{equation}
where $g(u)$ is a nonlinear function of $u$. Eq.(\ref{k-g-eq}) is
used for the description of particle dynamics in relativistic
quantum mechanics and includes a great number of PDEs in
mathematical physics. For a cubic nonlinearity $g(u)=u^3-u$, it is
used as a model in field theory \cite{dashen-1974}.
Eq.(\ref{k-g-eq}) with a $\sin u$ term is named by sine-Gordon
equation which has various applications and can be solved by inverse
scattering method \cite{sine}.
Eq.(\ref{k-g-eq}) also includes sinh-Gordon equation with $g(u)=e^u
\pm e^{-u}$, Tzetzeica equation with $g(u)=e^u \pm e^{-2u}$ and
Mikhailov equation  with $g(u)=e^{2u} \pm e^{-u}$, which are all
soliton equations.

Conservation laws of Eq.(\ref{k-g-eq}) had been studied by multiplier method and variational symmetry method in \cite{sb-2002a,Olver}
respectively. In particular, since there exists a Lagrangian
$L=u_{x}^2/2-u_{t}^2/2-\int g(u)du$ for
Eq.(\ref{k-g-eq}), thus the determining equations for the adjoint
symmetry and symmetry are identical. Moreover, there is a one-to-one correspondence between multipliers and conservation laws for Eq.(\ref{k-g-eq}).

\subsubsection{Nonlinear self-adjointness with differential
substitution} In order to demonstrate the effectiveness of Theorem
\ref{th-2}, we start with the equality (\ref{equ-id1-id2}) to show
that Eq.(\ref{k-g-eq}) is nonlinearly self-adjoint with differential
substitution. 

Let the formal Lagrangian of Eq.(\ref{k-g-eq})
\begin{eqnarray}
\mathcal {L}=\alpha(u_{tt}-u_{xx}-g(u))
\end{eqnarray}
with a new introduced dependent variable $\alpha$, then the adjoint
equation of Eq.(\ref{k-g-eq}) is
\begin{eqnarray}\label{adjoint}
&& \frac{\delta \mathcal {L}}{\delta
u}=D^2_t\alpha-D^2_{x}\alpha-g'(u)\alpha.
\end{eqnarray}

Assume the differential substitution
$\alpha=\varphi(x,t,u,\partial_xu,\partial_tu,\dots,\partial^p_xu,\partial^{p-1}_x\partial_tu)$
and use the equality (\ref{equ-id1-id2}), then one has
\begin{eqnarray}\label{det}
&&\no\mathscr{D}^2_t\varphi
+\mathscr{R}_\varphi[u]G+\sum_{i=0}^{p-1}D_x^iG\left[D_t(\varphi_{\partial_x^i\partial_tu})+\mathscr{D}_t(\varphi_{\partial_x^i\partial_tu})\right]
-D^2_{x}\varphi-g'(u)\varphi\\
&&\hspace{3cm}=\sum_{j=0}^p
\lambda_jD_x^jG+\sum_{k=0}^{p-1}\mu_kD_x^kD_tG+\sum_{l,s=0}^{p-1}\nu_{ls}D_x^lG\,D_x^sG,
\end{eqnarray}
where, hereinafter,
$\lambda_j,\mu_k,\nu_{ls}(i,k,l,s=1,\dots,p-1)$ are arbitrary
functions of $x,t,u$ and up to $p+2$ order derivatives of $u$
without containing $u_{tt}$ and its differential results, and
$$\mathscr{D}_t=\partial_t+u_t\partial_u+u_{xt}\partial_{u_x}+(u_{xx}+g(u))\partial_{u_t}
+\dots$$ is the total derivative operator which expresses $u_{tt}$
and its derivatives through Eq.(\ref{k-g-eq}). In particular, $D_t\varphi=\mathscr{D}_t\varphi+\mathscr{R}_\varphi[\partial_tu]G$ and
$D_t =\mathscr{D}_t$ on the solution space of  Eq.(\ref{k-g-eq}). Note that the differential substitution $\varphi$ does not involve
$u_{tt}$ and its differential results since they can be eliminated
by Eq.(\ref{k-g-eq}).

By considering whether the terms in Eq.(\ref{det}) contain $u_{tt}$
and its differential consequences or not, we obtain the determining
system of the substitution $\varphi$, consisting of
\begin{eqnarray}\label{symm-con}
&&\mathscr{D}^2_t\varphi-D^2_{x}\varphi-g'(u)\varphi=0,
\end{eqnarray}
which is the determining system for a symmetry
$X=\varphi\,\partial_u$ of Eq.(\ref{k-g-eq}), and an extra
determining condition on $\varphi$
\begin{eqnarray}\label{extra-con}
&&\no \mathscr{R}_\varphi[u]G+\sum_{i=0}^{p-1}D_x^iG\left[D_t(\varphi_{\partial_x^i\partial_tu})+\mathscr{D}_t(\varphi_{\partial_x^i\partial_tu})\right]\\
&&\hspace{3cm}=\sum_{j=0}^p
\lambda_jD_x^jG+\sum_{k=0}^{p-1}\mu_kD_x^kD_tG+\sum_{l,s=0}^{p-1}\nu_{ls}D_x^lGD_x^sG.
\end{eqnarray}

Since $D_t\Delta=\mathscr{D}_t\Delta+\mathscr{R}_\Delta[\partial_tu]G$
for the function $\Delta$, thus Eq.(\ref{extra-con})
becomes
\begin{eqnarray}\label{extra-con-beco}
&&\no\mathscr{R}_\varphi[u]G+\sum_{i=0}^{p-1}D_x^iG\left[\mathscr{R}_{\varphi_{\partial_x^i\partial_tu}}[\partial_tu]G
+2\mathscr{D}_t(\varphi_{\partial_x^i\partial_tu})\right]
\\
&&\hspace{3cm}=\sum_{j=0}^p
\lambda_jD_x^jG+\sum_{k=0}^{p-1}\mu_kD_x^kD_tG+\sum_{l,s=0}^{p-1}\nu_{ls}D_x^lGD_x^sG.
\end{eqnarray}
Then splitting  Eq.(\ref{extra-con-beco}) with respect to $G$ and
its differential consequents, we obtain
\begin{eqnarray}\label{extra-con-beco-3}
&&\no \varphi_{\partial_x^k\partial_tu}-\mu_k=0,\\
&&\no\varphi_{\partial_x^i\partial_tu~\partial_x^j\partial_tu}-\nu_{ij}=0,\\
&&\no\varphi_{\partial_x^iu}+2\mathscr{D}_t(\varphi_{\partial_x^i\partial_tu})
-\lambda_i=0,\\
&&\varphi_{\partial_x^pu}-\lambda_p=0,~~~i,j,k=0,1,\dots,p-1.
\end{eqnarray}

Since $\lambda_i,\mu_k,\nu_{ij}$ and $\lambda_p$ are undetermined
functions of their arguments, thus system (\ref{extra-con-beco-3})
holds identically which demonstrates that the essential requirement
of a function $\varphi$ to be a differential substitution of
Eq.(\ref{k-g-eq}) is the symmetry determining system
(\ref{symm-con}). Then we have:
\begin{theorem}\label{th-1}
The characteristic
$\varphi=\varphi(x,t,u,\partial_xu,\partial_tu,\dots,\partial^p_xu,\partial^{p-1}_x\partial_tu)$
of a symmetry $X=\varphi\,\partial_u$ is a differential substitution
of nonlinear self-adjointness and vice versa.
\end{theorem}

Theorem \ref{th-1} means that finding differential substitution is turned into solve symmetry determining system (\ref{symm-con}). On the other hand, since Eq.(\ref{k-g-eq}) comes from a Lagrangian and possesses conservation laws for energy, momentum, etc, it is automatically nonlinearly self-adjoint. Moreover, since the adjoint-symmetries are same as symmetries for Eq.(\ref{k-g-eq}), nonlinear self-adjointness with differential substitution is equivalent to the existence of generalized symmetries.
\subsubsection{Relation to multiplier method}
 We use multiplier method to study conservation law of
Eq.(\ref{k-g-eq}) in order to compare it with nonlinear
self-adjointness with differential substitution.

Following the idea of multiplier method \cite{Olver,blu11,sb-2002b},
a function $\Lambda=\Lambda(x,t,u,\partial_xu,\partial_tu,\dots,$
$\partial^p_xu,\partial^{p-1}_x\partial_tu)$ is a multiplier of
Eq.(\ref{k-g-eq}) if and only if Euler operator annihilates $\Lambda
G$ identically, that is
\begin{eqnarray}\label{multi-con}
&& \no\frac{\delta (\Lambda G)}{\delta
u}=D^2_t\Lambda-D^2_{x}\Lambda-g'(u)\Lambda+\Lambda_u G\\\no
&&\hspace{1.8cm}-D_x(\Lambda_{\partial_x
u}G)+\dots+(-D_x)^p(\Lambda_{\partial_x^p
u}G)+(-D_x)^{p-1}D_t(\Lambda_{\partial_x^{p-1}\partial_t u}G)\\
&&\hspace{1.35cm}=0.
\end{eqnarray}

On the solution space of Eq.(\ref{k-g-eq}), collecting the separation of Eq.(\ref{multi-con}) in terms of $G$
and its differential consequents yields a determining system for the
multiplier $\Lambda$, which contains the symmetry determining
equation
\begin{eqnarray}\label{sym-det-1}
&&\mathscr{D}^2_t\Lambda-D^2_{x}\Lambda-g'(u)\Lambda=0,
\end{eqnarray}
and the so-called ``adjoint invariance conditions" or `` Helmholtz-type conditions"
\begin{eqnarray}\label{extra-con-beco-4}
&&\no(1+(-1)^p)\Lambda_{\partial_x^pu}=0,\\
&&\no \Lambda_{\partial_x^i\partial_tu}+\sum_{j=0}^{p-1}(-1)^{j+1}{j\choose i} D_x^{j-i}\Lambda_{\partial_x^j\partial_t u}=0,\\
&&\Lambda_{\partial_x^i\partial_tu~\partial_x^k\partial_tu}
+\sum_{j=0}^{p-1}(-1)^{j+1}{j\choose i} D_x^{j-i}\Lambda_{\partial_x^j\partial_t u~\partial_x^k\partial_tu}=0,~~k=1,\dots,p-1,\\
&&\no\Lambda_{\partial_x^iu}+2\mathscr{D}_t\Lambda_{\partial_x^i\partial_tu}
+\sum_{j=0}^{p-1}(-1)^{j+1}{j\choose
i}D_x^{j-i}\mathscr{D}_t\Lambda_{\partial_x^j\partial_t u}
+\sum_{j=0}^{p}(-1)^j{j\choose i}D_x^{j-i}\Lambda_{\partial_x^j
u}=0,
\end{eqnarray}
where ${j\choose i}=j!/(i!(j-i)!)$.

Summarizing the above computations, we obtain:
\begin{theorem}\label{th-21}
A function
$\Lambda=\Lambda(x,t,u,\partial_xu,\partial_tu,\dots,\partial^p_xu,\partial^{p-1}_x\partial_tu)$
is a multiplier of Eq.(\ref{k-g-eq}) if and only if it
satisfy system (\ref{sym-det-1}) and (\ref{extra-con-beco-4}).
\end{theorem}

Obviously, the conditions of multiplier for Eq.(\ref{k-g-eq}) are
system (\ref{sym-det-1}) and (\ref{extra-con-beco-4}) while the
condition of differential substitution is only (\ref{sym-det-1}),
thus the set of differential substitutions includes the one of
multipliers as a subset.

\subsubsection{Conservation law}
By Theorem \ref{con-law}, a general conservation law formula of
Eq.(\ref{k-g-eq}) is given as follows.
\begin{theorem}\label{th-31}
Assume a symmetry
$X=\eta(x,t,u,\partial_xu,\partial_tu,\dots)\partial_u$ leaves
Eq.(\ref{k-g-eq}) invariant, then a conservation law
\begin{eqnarray}
\no &&D_t C^t+D_x C^x=\big(\mathscr{W}_{\mathscr{D}_t\eta}[\partial_tu]\alpha-\mathscr{W}_{\eta}[\partial_tu]\mathscr{D}_t\alpha
+\mathscr{W}_\alpha[\partial_tu]\mathscr{D}_t\eta-\mathscr{W}_{\mathscr{D}_t\alpha}[\partial_tu]\eta\big)*G
\end{eqnarray} is given by the formulae
\begin{eqnarray}\label{con-tele}
&& C^t=\alpha \mathscr{D}_t \eta-\eta \mathscr{D}_t\alpha,~~ C^x=\eta D_x\alpha-\alpha
D_x \eta,
\end{eqnarray}
where the differential substitution
$\alpha=\varphi(x,t,u,\partial_xu,\partial_tu,\dots,\partial^p_xu,\partial^{p-1}_x\partial_tu)(\neq\eta)$
is given by Theorem \ref{th-1}.
\end{theorem}

Given a differential substitution provided in Theorem \ref{th-1},
formulae (\ref{con-tele}) build a connection between symmetries and
conservation laws for Eq.(\ref{k-g-eq}). 

In what follows, we apply Theorem \ref{th-31}  to construct local
conservation laws of Klein-Gordon equation (\ref{k-g-eq}).

For arbitrary function $g(u)$, Eq.(\ref{k-g-eq}) admits space translation symmetry $X_1=u_x\partial_u$, time translation symmetry
$X_2=u_t\partial_u$ and rotation symmetry $X_3=(tu_x+xu_t)\partial_u$, then by Theorem \ref{th-1}
we obtain three differential substitutions
\begin{eqnarray}
&&\varphi_1=u_x,~\varphi_2=u_t,~\varphi_3=tu_x+xu_t.
\end{eqnarray}

A symmetry $X_\eta=\eta^i(x,u,u_{(1)},\dots,u_{(s)})\partial_{u^i}$ of
Eq.(\ref{perturb}) is a variational symmetry if and only if
$\mbox{pr}X_\eta (L)+\sum_{i=1}^nD_i(\xi_i\,L)=\mbox{Div}(B)$ holds
for all $x,u$, where $L$ is the Lagrangian of Eq.(\ref{perturb}),
$\mbox{pr} X_\eta(L)$ denotes the proper prolongation of $X_\eta$
and $\mbox{Div}(B)$ is the divergence expression for some
$B=(B_1,\dots,B_n)$ \cite{Olver,blu11}.

It is easy to show that symmetries $X_1,X_2$ and $X_3$ are
variational symmetries and by Noether' theorem corresponds to three
conservation laws $\big(D_t\widetilde{C}_i^t+D_x\widetilde{C}_i^x\big)_{|G=0}=0\,
(i=1,2,3)$ where the conserved currents are given by
\begin{eqnarray}\label{con-law-noether}
&&\no \widetilde{C}_1^t=-u_xu_t,~~
\widetilde{C}_1^x=\frac{1}{2}(u_x^2+u_{t}^2)+\int g(u)du;\\
&&\no \widetilde{C}_2^t=\frac{1}{2}(u_x^2+u_{t}^2)-\int g(u)du,~~
\widetilde{C}_2^x=-u_xu_t;\\\no
&&\widetilde{C}_3^t=\frac{1}{2}x\Big(u_{x}^2-3u_t^2-2\int
g(u)du\Big)-tu_xu_t ,\\
&&\widetilde{C}_3^x=\frac{1}{2} t\Big(3u_{x}^2-u_t^2-2\int
g(u)du\Big)+xu_xu_t.
\end{eqnarray}

On the other hand, from the point of view of nonlinear
self-adjointness with differential substitution and by Theorem
\ref{th-31}, the conservation law associated
with $\varphi_1$ $$D_tC_1^t+D_xC_1^x=\big(\mathscr{W}_{\mathscr{D}_t\eta}[\partial_tu]u_x-\mathscr{W}_{\eta}[\partial_tu]u_{xt}
+D_x\eta\big)*G$$ is given by $C_1^t=u_x \mathscr{D}_t \eta-\eta u_{xt}$ and $C_1^x=\eta u_{xx}-u_x D_x\eta$,
while the one associated with $\varphi_2$
$$D_tC_2^t+D_xC_2^x=\big[\mathscr{W}_{\mathscr{D}_t\eta}[\partial_tu]u_t-\mathscr{W}_{\eta}[\partial_tu](u_{xx}+g(u))+\mathscr{D}_t\eta\big]*G$$ is determined by $C_2^t=u_t \mathscr{D}_t \eta-\eta u_{xx}-\eta g(u)$ and $ C_2^x=\eta u_{tx}-u_tD_x \eta$,
and the one associated with $\varphi_3$
\begin{eqnarray}
&&\no D_tC_3^t+D_xC_3^x=\big[\mathscr{W}_{\mathscr{D}_t\eta}[\partial_tu](tu_x+xu_t)+tD_x\eta\\
&&\no\hspace{3.2cm}-\mathscr{W}_{\eta}[\partial_tu](xu_{xx}+tu_{xt}+xg(u)+u_x)
+x\mathscr{D}_t\big]*G
\end{eqnarray}
is expressed by
\begin{eqnarray}\label{con-tele-3}
&&\no  C_3^t=(tu_x+xu_t) \mathscr{D}_t \eta- (xu_{xx}+x
g(u)+tu_{xt}+u_x)\eta,\\&&\no C_3^x=(u_t+tu_{xx}+xu_{xt})\eta-
(tu_x+xu_t)D_x \eta,
\end{eqnarray}
where  $X=\eta\partial_u$ is a symmetry of Eq.(\ref{k-g-eq}).

Observe the above two methods for Eq.(\ref{k-g-eq}) with arbitrary
$g(u)$, we find that Noether' theorem constructs first-order local
conservation laws determined by (\ref{con-law-noether}) while
nonlinear self-adjointness with differential substitution method
generates high-order local and nonlocal conservation laws, where
nonlocal ones arise from nonlocal differential substitutions. Moreover, since any of the obtained conservation laws corresponds to a multiplier, multiplier method yields all conservation laws of  Eq.(\ref{k-g-eq}) while formula (\ref{con-tele}) only gives parts of them such as high-order local and nonlocal ones.

We consider a special case $g(u)=u^n$, which corresponds to
the Klein-Gordon equation with power law nonlinearity
\begin{eqnarray}\label{power}
&&G_{kg}=u_{tt}-u_{xx}-u^n=0, ~~~~~n\neq -1,0,1,
\end{eqnarray}
which is studied in the context of relativistic quantum mechanics.
The Lie point symmetries admitted by Eq.(\ref{power}) are extended
by $X_4=\big(2u/(n-1)+xu_x+tu_t\big)\partial_u$ in addition to $X_i(i=1,2,3)$.

With the help of the criterion of variational symmetry
\cite{Olver}, we find that the expression
\begin{eqnarray}
&& \no\mbox{pr} X_4(L)+D_x(-x \,L)+D_t(-t\, L)= \frac{2}{3
(n-1)}\big(3 u_x^2-3 u_t^2+nu^3-4 u^3\big),
\end{eqnarray}
does not take the divergence form for any $x,t$ and $u$, where
$L=u_{x}^2/2-u_{t}^2/2-u^{n+1}/(n+1)$ is
the Lagrangian of Eq.(\ref{power}) and $\mbox{pr} X_4(L)$ denotes
the first-order prolongation of $X_4$, thus $X_4$ is not a
variational symmetry and cannot be used to construct conservation
law via Noether' theorem. However, in the context of nonlinear
self-adjointness with differential substitution, one can use it to
generate conservation laws of Eq.(\ref{power}) from the following
two aspects:

(I). For example, substitute $\eta=2u/(n-1)+xu_x+tu_t$ and $\alpha=u_x$ into
formula (\ref{con-tele}), one obtains a nontrivial conservation law of Eq.(\ref{power})
 $$D_t \widehat{C}_{(\ref{power})}^t+D_x
\widehat{C}_{(\ref{power})}^x=\frac{n+3}{n-1}u_x*G_{kg},$$
where $\Lambda=(n+3)/(n-1)u_x$ is a multiplier and
 \begin{eqnarray}
&&\no  \widehat{C}_{(\ref{power})}^t= t u_x u^n-tu_t u_{xt}+\frac{1}{n-1}\big(n\,u_xu_t+ u_xu_t-2 u u_{xt}\big),\\
&&\no \widehat{C}_{(\ref{power})}^x=t u_tu_{xx} +\frac{1}{2 (n-1)}\big(4
uu_{xx}-n\,u_x^2-3 u_x^2\big).
\end{eqnarray}

(II). On the other hand, by Theorem \ref{th-1}, the characteristic
of symmetry $X_4$ is a differential substitution of nonlinear
self-adjointness, i.e., $\alpha=2u/(n-1)+xu_x+tu_t$. Then by
Theorem \ref{th-31}, a nontrivial conservation law
\begin{eqnarray}
&& \no D_t \widetilde{C}_{(\ref{power})}^t+D_x
\widetilde{C}_{(\ref{power})}^x
=\Big(\mathscr{W}_{\mathscr{D}_t\eta}[\partial_tu]\alpha-\mathscr{W}_{\eta}[\partial_tu]\mathscr{D}_t\alpha-\frac{2\eta}{n-1}+x D_x\eta+t \mathscr{D}_t\eta\Big)*G_{kg}
\end{eqnarray}
 is given by the formulae
\begin{eqnarray}
&&\no \widetilde{C}_{(\ref{power})}^t=\Big(\frac{2u}{n-1}+xu_x+tu_t\Big) \mathscr{D}_t \eta- \Big(\frac{2u_t}{n-1}+xu_{xt}+u_t+tu_{xx}+tu^n \Big)\eta,\\
&&\no
\widetilde{C}_{(\ref{power})}^x=\Big(\frac{2u_x}{n-1}+u_x+xu_{xx}+tu_{xt}\Big)\eta-\Big(\frac{2u}{n-1}+xu_x+tu_t\Big)
D_x \eta,
\end{eqnarray}
where $X=\eta\,\partial_u$ is a symmetry admitted by
Eq.(\ref{power}) such that the multiplier is not zero.

\section{Conclusion}
We show that the set of adjoint symmetries admitted by the PDEs is
identical to the one of differential substitutions of nonlinear
self-adjointness, and then express the correspondence between
symmetries, adjoint symmetries and conservation laws via formula
(\ref{formula}), which avoids integral operation by multiplier
method. 
Furthermore, we demonstrate that the set of differential substitution of nonlinear self-adjointness contains the
one of conservation law multipliers as a subset. Three different
types of examples illustrate our results.
In addition, the presented results, after proper arrangements, can
be applied to study approximate nonlinear self-adjointness of
perturbed PDEs \cite{ib-2011,zhang-2013,zhang-2014}.

\section*{Acknowledgments}
We sincerely appreciate the referees for valuable
comments and improvements. This paper is supported by the National Natural Science Foundation
of China (Nos. 11671014 and 11301012), Beijing Natural Science Foundation
(No.1173009), Scientific Research Project of Beijing Educational Committee (No.KM201710009011).

\end{document}